\documentclass[
aps,nofootinbib,
showpacs,showkeys,preprint
tightenlines,preprintnumbers,] {revtex4}

\usepackage{epsf,epsfig,subfigure,graphicx,amsmath,amssymb 
}
\usepackage{color}
\newcommand{\dis}[1]{\begin{equation}\begin{split}#1\end{split}\end{equation}}
\newcommand{\ie}{{\it i.e.~}}
\newcommand{\etal}{{\it et al.\,}}

\newcommand{\Qem}{Q_{\rm em}}

\newcommand{\gev}{\,\textrm{GeV}}

\newcommand{\Mpt}{$M_{\rm P}$}

\newcommand{\Mgt}{$M_{\rm GUT}$}

\newcommand{\Umutau}{U(1)$_{\mu-\tau}$}

\def\sw0{{$\sin^2\theta_W^0$}}

\def\E6{{\rm E_6}}

\def\EE8{{\rm E_8\times E_8'}}

\def\one{{\bf 1}}

\begin{document}

\draft

\title{\Large\bf   $L_{\mu}-L_{\tau}$ effects to quarks and leptons from flavor unification
}

\author{ Paul H. Frampton$^{(1)}$, 
Sin Kyu Kang$^{(2)}$, Jihn E.  Kim$^{(3,4)}$, Soonkeon Nam$^{(3)}$ }

\address
{ $^{(1)}$ Dipartimento di Matematica e Fisica ``Ennio De Giorgi'',
Universita del Salento and INFN-Lecce,
Via Arnesano, 
73100 Lecce, Italy,  \\
$^{(2)}$
School of Liberal Arts, Seoul University of Science and Technology, Seoul 01811, Republic of Korea,  \\
$^{(3)}$Department of Physics, Kyung Hee University, 26 Gyungheedaero, Dongdaemun-Gu, Seoul 02447, Republic of Korea,  \\
 $^{(4)}$ Department of Physics and Astronomy, Seoul National University, 1 Gwanakro, Gwanak-Gu, Seoul 08826, Republic of Korea
}

\begin{abstract} 
\noindent
In the family grand unification models (fGUTs), we propose that gauge U(1)'s beyond the minimal GUT gauge group are family gauge symmetries.  For the symmetry $L_\mu-L_\tau$, \ie $Q_{2}-Q_{3}$ in our case, to be useful for the LHC anomaly, we discuss an SU(9) fGUT and also present an example in  Georgi's SU(11) fGUT.
\end{abstract}
\maketitle


\section{Introduction}\label{sec:Introduction}

\noindent
The most interesting problem remaining in the Standard Model (SM) is the flavor or family problem. The family problem forces on the symmetry of all the massless chiral fields surviving below the grand unification (GUT) scale \Mgt\, or  the Planck scale \Mpt. Chiral fields with quantum numbers consistent with the observed weak and electromagnetic phenomena were a crucial achievement of the SM. With the electromagnetic and charged currents (CCs), the leptons need representations which are a doublet or bigger. A left-handed (L-handed) lepton doublet  $(\nu_e,e)$ alone is not free of gauge anomalies because the observed electromagnetic charges are not $\pm \frac12$. The anomalies from the fractional electromagnetic charges of the $u$ and $d$ quarks add up to make the total anomaly from the first family vanish \cite{Bouchiat72,Jackiw72}. It is very difficult to obtain another kind of chiral model free of gauge anomalies. If another chiral model is found consistently with some observed fact, that model should include some truth in it. The same gauge structures of the first family, $\{\nu_e,e,u,d \}$, repeats two more times in the $\mu$ family and $\tau$ family.

\bigskip
\noindent
A correct treatment of flavors is necessary not only in the familiar field of particle theory and high-energy
physics but also in astronomy and especially cosmology. Big Bang Nucleosynthesis calculations can
place upper limits on the number of active neutrinos. The mixings of six flavors of quark allow a CP
violating phase which is successful in agreeing precisely with data on CP violation in K and B decays, and yet the correct
derivation of baryogenesis which itsef needs CP violation and the tiny ratio $\eta = (\Delta n_B / n_{\gamma}) \simeq 9 \times 10^{-11}$
remains challenging, especially whether the CP violation known in quark flavor mixing can suffice to explain the matter-antimatter asymmetry of the universe. These are merely
two examples of cosmolological applications of flavor theory.

\bigskip
\noindent
Recent phenomenological studies on the flavor problem centered around (i) ``Why are there three families?'', and (ii) ``What are the symmetries giving the observed Cabibbo--Kobayashi--Maskawa (CKM)  and the Pontecorvo--Maki--Nakagawa--Skata (PMNS) matrices?''.  
The first part of the family problem was formulated in simple gauge groups 40 years ago by Georgi \cite{Georgi79}. Some interesting models appeared along this line in \cite{Kim80,Frampton80}. The second part is discussed recently in 
\cite{MB19,FK19,Kimetal20,KimKim20} in relation to the CKM and PMNS matrices. Until recently, there has not appeared any significant deviation from the CKM and PMNS matrices. If some deviation were to be observed, then it might predict beyond the SM (BSM) physics, probably in the fourth family, sterile neutrinos, or in the important scalar interactions as presented in this paper. 

\bigskip
\noindent
The obvious family dependences are in the masses and the CKM and PMNS matrices.  Given these, the next level is to check lepton family universality  in the decays of mesons. With hundreds of millions of $B$ decays already found at the LHC, it is possible to check the universality of a  ratio of the type
\dis{
R_H=\frac{{\cal B}(B\to H\mu^+\mu^-)}{{\cal B}(B\to He^+e^-)}
}
where $H$ represents a hadron. In the last few years, observation of an anomaly in $R_H$ with a 2.5 $\sigma$ level significance \cite{LHCbRK14,LHCbRK19} (see also \cite{Belle19}) attracted a great deal of attention \cite{Ali00,Beneke01,Geng01,Ali06,Egede08,Bobeth08,Buras09}. In the leptonic sector, the lepton family dependence was used to make the BNL $(g-2)_\mu$ observation to draw near to the SM prediction. This needed an additional interaction for the muon family. If it is a gauge U(1) interaction, cancellation of anomalies necessitates to have contributions from  other family members, for example in the form of the quantum number such as $L_\mu-L_\tau$.

\bigskip
\noindent
Within  fGUT models,  $L_{\mu}-L_{\tau}$ symmetry can affect the prediction of $R_H$ phenomenology since the quantum number  $L_{\mu}-L_{\tau}$ applies also to the quark members in the same family.

\section{ ${\bf f}$GUT grand unification of of families}
 
\bigskip

\noindent
Let us use the tensor notation such that the index $A=\{1,2,\cdots,N\}$ is split into the GUT index $\alpha=\{1,2,\cdots,5\}$, family indices $I=\{6=$electron family,  $7=$muon family,  $8=$tau family\}, and $\{9,\cdots,N\}$ for dummy numbers. We will use only the completely antisymmetric representations so that higher dimensional quarks such as ${\bf 6, 6^*},$ etc. do not appear. 

\bigskip
\noindent
Consider the three family indices, $I=6,7,8$, for U(3) representations,  which are equivalently used as $I\to \{e\equiv 1, \mu\equiv 2, \tau\equiv 3 \}$. The upper ($X^{IJ\cdots}$) and lower indices ($X_{IJ\cdots}$)  are distinguished. The Levi-Civita symbols, $\epsilon^{IJK}$ and $\epsilon_{IJK}$ for the  U(3) group, will be used to raise and lower the indices.  Thus, $X^{(0,1,1)}=X_{(1,0,0)}$, etc. Note also that   complex conjugation corresponds to taking the hermitian conjugate, $X^{I*}=X_{-I}$.

\bigskip
\noindent
Now, we consider the following completely antisymmetric representations which split into
  \dis{
\Psi^A &=\psi^\alpha\oplus \one^{(1,0,0)}\oplus \one^{(0,1,0)}\oplus \one^{(0,0,1)}\oplus \one^{(0,0,0)}\oplus \cdots \\
\Psi^{AB} &=\psi^{\alpha\beta(0,0,0)} \oplus \psi^{\alpha(1,0,0)} \oplus \psi^{\alpha(0,1,0)} \oplus \psi^{\alpha(0,0,1)} \oplus \psi^{\alpha(0,0,0)} \oplus \cdots \\
&~ \oplus \one^{(1,1,0)} \oplus \one^{(1,0,1)} \oplus \one^{(0,1,1)} \oplus \one^{(1,0,0)}\oplus \one^{(0,1,0)}\oplus \one^{(0,0,1)}\oplus \one^{(0,0,0)}\oplus\cdots\\
\Psi^{ABC}& =\psi^{\alpha\beta\gamma}\oplus \psi^{\alpha\beta(1,0,0)}  \oplus \psi^{\alpha\beta(0,1,0)}  \oplus \psi^{\alpha\beta(0,0,1)}  \oplus \psi^{\alpha\beta(0,0,0)}   \oplus \cdots \\
&~\oplus \psi^{\alpha(1,1,0)}   \oplus \psi^{\alpha(1,0,1)} \oplus \psi^{\alpha(0,1,1)}   \oplus \psi^{\alpha(1,0,0)}  \oplus \psi^{\alpha(0,1,0)}  \oplus \psi^{\alpha(0,0,1)}  \oplus \psi^{\alpha(0,0,0)} \oplus \cdots\\
&~\oplus \one^{(1,1,1)} \oplus  \one^{(1,1,0)} \oplus  \one^{(1,0,1)} \oplus  \one^{(1,0,0)}\oplus  \one^{(0,1,0)}\oplus  \one^{(0,0,1)}  \oplus  \one^{(0,0,0)}\oplus\cdots   \\
\Psi^{ABCD}&=\psi^{\alpha\beta\gamma\delta}\oplus \psi^{\alpha\beta\gamma(1,0,0)}\oplus\psi^{\alpha\beta\gamma(0,1,0)}\oplus \psi^{\alpha\beta\gamma(0,0,1)}\oplus    \psi^{\alpha\beta\gamma(0,0,0)}\cdots\\
&~ \oplus   \psi^{\alpha\beta(1,1,0)} \oplus \psi^{\alpha\beta(1,0,1)}\oplus \psi^{\alpha\beta(0,1,1)} \oplus \psi^{\alpha\beta (1,0,0)}\oplus \psi^{\alpha\beta (0,1,0)}\oplus \psi^{\alpha\beta (0,0,1)}\oplus \psi^{\alpha\beta (0,0,0)}\oplus \cdots\\
&~\oplus \psi^{\alpha (1,1,1)}\oplus  \psi^{\alpha(1,1,0)}\oplus  \psi^{\alpha (1,0,1)}\oplus   \psi^{\alpha(1,0,0)}\oplus   \psi^{\alpha (0,1,0)}\oplus \psi^{\alpha(0,0,1)}\oplus 
 \psi^{\alpha (0,0,0)}\\
 &~\oplus \one^{(1,1,1)} \oplus \one^{(1,1,0)} \oplus \one^{(1,0,1)} \oplus \one^{(0,1,1)}  \oplus \one^{(1,0,0)} \oplus \one^{(0,1,0)} \oplus \one^{(0,0,1)} \\
&~\oplus~\textrm{etc.}
}

\bigskip
\noindent
The example for SU(9) given in \cite{Frampton80} thus becomes, after removing vectorlike representations,
\begin{eqnarray}
\Psi^{ABC}\oplus 9\Psi_A=&& \psi_{\alpha\beta}\oplus \psi^{\alpha\beta(1,0,0)}  \oplus \psi^{\alpha\beta(0,1,0)}  \oplus \psi^{\alpha\beta(0,0,1)}  \oplus \psi^{\alpha\beta(0,0,0)} \nonumber    \\
&& \oplus \psi^{\alpha(1,1,0)}   \oplus \psi^{\alpha(1,0,1)} \oplus \psi^{\alpha(0,1,1)} \oplus \psi^{\alpha(1,0,0)}   \oplus \psi^{\alpha(0,1,0)} \oplus \psi^{\alpha(0,0,1)} \nonumber  \\
&& \oplus \one^{(1,1,1)}\oplus \one^{(1,1,0)}\oplus \one^{(1,0,1)}\oplus \one^{(0,1,1)} \nonumber   \\
 && \oplus 9\psi_\alpha\oplus 9\cdot \one_{(-1,0,0)}\oplus 9\cdot \one_{(0,-1,0)}\oplus 9\cdot \one_{(0,0,-1)}\oplus 9\cdot \one^{(0,0,0)}\nonumber  \\
 =&&  \psi^{\alpha\beta(1,0,0)}  \oplus \psi^{\alpha\beta(0,1,0)}  \oplus \psi^{\alpha\beta(0,0,1)}\nonumber    \\
&&\oplus \psi^{\alpha(1,1,0)}   \oplus \psi^{\alpha(1,0,1)} \oplus \psi^{\alpha(0,1,1)} \oplus \psi^{\alpha(1,0,0)}   \oplus \psi^{\alpha(0,1,0)} \oplus \psi^{\alpha(0,0,1)} \oplus 9\psi_{\alpha  (0,0,0)}\label{eq:SU9}\\
&&\oplus 8\cdot \one_{(-1,0,0)}\oplus 8\cdot \one_{(0,-1,0)}\oplus 8\cdot \one_{(0,0,-1)}\oplus 10\cdot \one^{(0,0,0)}.\nonumber
\end{eqnarray}
The remaining three lepton (doublet) families, out of the second line from the bottom of Eq. (\ref{eq:SU9}), will be $3\, \psi_{\alpha(0,0,0)}$ which do not carry $L_{\mu}-L_\tau$.  
To avoid confusion, we will write quantum numbers of $L_{\mu}-L_\tau$ always as subscripts within square brackets.
 
\bigskip
\noindent
Georgi's fGUT model is for $N=11$ \cite{Georgi79},
\dis{
\Psi^{ABCD}+ \Psi_{ABC}+ \Psi_{AB}+ \Psi_{A}.
}
Defining the first three slots for
  \dis{
\rm U(1)_6\times U(1)_7\times U(1)_8,~{\rm or~renamed~as}~U(1)_1\times U(1)_2\times U(1)_3
}
we obtain
\dis{
\Psi^{ABCD}\to  
&\psi^{\alpha\beta\gamma\delta}\oplus  \psi^{\alpha\beta\gamma(1,0,0)}\oplus  \psi ^{\alpha\beta\gamma(0,1,0)}\oplus  \psi ^{\alpha\beta\gamma(0,0,1)}\oplus 3\cdot \psi^{\alpha\beta\gamma(0,0,0)} \\
&\oplus  \psi^{\alpha\beta (1,1,0)}\oplus\psi^{\alpha\beta (1,0,1)}\oplus   \psi^{\alpha\beta (0,1,1)}\oplus 3\cdot \psi^{\alpha\beta (1,0,0)}  \oplus  3\cdot \psi^{\alpha\beta (0,1,0)}\oplus 3\cdot \psi^{\alpha\beta (0,0,1)} \oplus 3\cdot  \psi^{\alpha\beta}_{(0,0,0)}\\
& \oplus   \psi^{\alpha (1,1,1)}\oplus 3\cdot \psi^{\alpha(1,1,0)}\oplus 3\cdot \psi^{\alpha(1,0,1)}   \oplus  3\cdot \psi^{\alpha(0,1,1)}\oplus 3\cdot \psi^{\alpha(1,0,0)}\oplus  3\cdot \psi^{\alpha(0,1,0)} \oplus  3\cdot \psi^{\alpha}_{(0,0, 1)}\oplus   \psi^{\alpha(0,0,0)} \\
&\oplus {3\cdot \one}^{(1,1,1)}\oplus 3\cdot {\one}^{(1,1,0)}\oplus 3\cdot {\one}^{(1,0,1)}\oplus 3\cdot {\one}^{(0,1,1)}\oplus    {\one}^{(1,0,0)}\oplus {\one}^{(0,1,0)}\oplus {\one}^{(0,0,1)} \oplus {\one}^{(0,0,0)}  \\
\Psi^{ABC}\to &  \psi^{\alpha\beta\gamma} \oplus  \psi^{\alpha\beta (1,0,0)}\oplus  \psi^{\alpha\beta(0,1,0)}\oplus  \psi^{\alpha\beta (0,0,1)}\oplus 3\cdot \psi^{\alpha\beta (0,0,0)} \\
&\oplus  \psi^{\alpha (1,1,0)}\oplus\psi^{\alpha(1,0,1)}\oplus   \psi^{\alpha (0,1,1)}\oplus 3\cdot \psi^{\alpha(1,0,0)}  \oplus  3\cdot \psi^{\alpha(0,1,0)}\oplus 3\cdot \psi^{\alpha(0,0,1)} \oplus 3\cdot  \psi^{\alpha(0,0,0)}\\
& \oplus    \one^{(1,1,1)}\oplus 3\cdot {\one}^{(1,1,0)}\oplus 3\cdot {\one}^{(1,0,1)}  \oplus 3\cdot {\one}^{(0,1,1)}   \oplus  3\cdot {\one}^{(1,0,0)}\oplus  3\cdot{\one}^{(0,1,0)}\oplus  3\cdot {\one}^{(0,0,1)} \oplus    {\one}^{(0,0,0)} \\
\Psi^{AB}\to &  \psi^{\alpha\beta}  \oplus  \psi^{\alpha(1,0,0)}\oplus  \psi^{\alpha(0,1,0)}\oplus  \psi^{\alpha(0,0,1)}\oplus 3\cdot \psi^{\alpha(0,0,0)} \\
&\oplus {\one}^{(1,1,0)}\oplus{\one}^{(1,0,1)}\oplus  {\one}^{(0,1,1)}\oplus 3\cdot {\one}^{(1,0,0)}  \oplus  3\cdot {\one}^{(0,1,0)}\oplus 3\cdot {\one}^{(0,0,1)} \oplus 3\cdot {\one}^{(0,0,0)}\\ 
\Psi^{A}\to &  \psi^{\alpha}  \oplus  {\one}^{(1,0,0)}\oplus {\one}^{(0,1,0)}\oplus  {\one}^{(0,0,1)}\oplus 3\cdot {\one}^{(0,0,0)}  
}
which leads to, removing vectorlike representations,  
\dis{
\Psi^{ABCD}& \oplus \Psi_{ABC} \oplus \Psi_{AB} \oplus \Psi_{A}=    3\cdot \psi_{\alpha\beta(0,0,0)}  \oplus\psi_{\alpha\beta}^{(1,0,0)}\oplus  \psi _{\alpha\beta}^{(0,1,0)}\oplus  \psi _{\alpha\beta}^{(0,0,1)}~~(*)\\
& \oplus 2\cdot \psi^{\alpha\beta (1,0,0)}  \oplus  2\cdot \psi^{\alpha\beta (0,1,0)}\oplus 2\cdot \psi^{\alpha\beta (0,0,1)}~~(**) \\
&\oplus      \psi^{\alpha\beta (1,0,0)}  \oplus    \psi^{\alpha\beta (0,1,0)}\oplus   \psi^{\alpha\beta (0,0,1)}  ~~(**)\\
&\oplus  \psi_{\alpha}^{ (-1,0,0)}\oplus\psi_{\alpha}^{ (0,-1,0)}\oplus   \psi_{\alpha}^{ (0,0,-1)}\\
& \oplus 6\cdot  \psi_{\alpha(0,0,0)}\oplus   \psi_{\alpha(-1,0,0)}  \oplus    \psi_{\alpha(0,-1,0)}\oplus   \psi_{\alpha(0,0,-1)}\\
&  \oplus   3\cdot \psi^{\alpha(1,0,0)} \oplus  3\cdot \psi^{\alpha (0,1,0)} \oplus  3\cdot \psi^{\alpha (0,0, 1)} \\
& \oplus 3\cdot {\one}^{(0,0,-1)}\oplus 3\cdot {\one}^{(0,-1,0)}  \oplus 3\cdot {\one}^{(-1,0,0)}   \oplus 4\cdot {\one}_{(-1,0,0)}  \oplus  4\cdot {\one}_{(0,-1,0)}\oplus 4\cdot {\one}_{(0,0,-1)}   .\label{eq:light}
}
  
 \bigskip
 \noindent
 We  want to define family numbers such that electron carries electron number, muon carries muon number and tau carries tau number.  Let $S_i$ be SU(5)-singlet scalars.

 \subsection{Notation for $L_\mu-L_\tau$}\label{subsec:LmuLtau}
 \noindent
 Since $\psi_\alpha$ contains lepton doublets, it is better to define lepton family number by $\psi_{AB}\to \psi_{\alpha\,I}$. So, the lepton family number is defined by the subscript $I$: electron doublet from $\psi_{\alpha\,6}$, muon   doublet from $\psi_{\alpha\,7}$, and tau   doublet from $\psi_{\alpha\,8}$. For the  $L^-\equiv L_\mu-L_\tau$ quantum numbers, we have  $\psi_{\alpha[0]}$ for electron family,  $\psi_{\alpha[1]}$ for muon family,  and $\psi_{\alpha[-1]}$ for tau family. We can use the U(3) Levi-Cevita symbol to raise or lower the family indices. For example, $\psi^{\alpha\beta IJ}$ is  $\psi^{\alpha\beta}_ K$ where $K$ is different from $I$ and $J$. Then, $S_e^{(1,0,0)}$ is $S_{e[0]}$, etc.  Let $L^+$ be the $L_\mu+L_\tau$ quantum number.  $S_{e[0]}$ does not break $L^+$ and $L^-$. But, the VEV  of $S_1^{(0,1,1)}$  breaks U(1)$_\mu\times$U(1)$_\tau$ to U(1)$_{\mu-\tau}$.  
 With $S_e$ and $S_1$, 
 the following couplings combine some vectorlike pairs  in the (**) lines of Eq. (\ref{eq:light}),
 \dis{
 \psi_{\alpha\beta}^{(0,1,0)} S_e^{(1,0,0)} \psi^{\alpha\beta(0,0,1)},~\psi_{\alpha\beta}^{(0,0,1)} S_e^{(1,0,0)} \psi^{\alpha\beta(0,1,0)},~\psi_{\alpha\beta}^{(0,0,0)} S_{1}^{(0,1,1)} \psi^{\alpha\beta(1,0,0)}.\nonumber
 }
 Therefore, from  the (**) lines  in Eq. (\ref{eq:light})  there remain, after the \Umutau~preserving VEVs of $\langle  S_e^{(1,0,0)}\rangle$ and $\langle  S_1^{(0,1,1)}\rangle,$
 \dis{
  \psi_{\alpha\beta}^{(1,0,0)}\oplus 2\psi^{\alpha\beta(0,1,0)}\oplus 2\psi^{\alpha\beta(0,0,1)},~
 }
 which gives three families and one vectorlike pair. With the exact $L_\mu-L_\tau$ conservation, there is no way to remove the remaining vectorlike pair. We must break the gauge symmetry U(1)$_{\mu-\tau}$ to remove the vectorlike pair. Let us introduce $S_{\mu}^{(0,1,0)}$ such that $\langle S_{\mu}\rangle\ll \langle S_{e}\rangle, \langle S_{1}\rangle$. The \Umutau~breaking effects to low energy physics is subdominant compared to the \Umutau~preserving interactions.  
 The  coupling 
 \dis{
 \psi_{\alpha\beta}^{(1,0,0)} S_\mu^{(0,1, 0)}\psi^{\alpha\beta(0,0,1)}
 }
removes the vectorlike pair.
  Thus, finally we obtain the following  three {\bf 10}'s,\footnote{If we used $S_\tau^{(-1,-1,2)}$ instead of $S_\mu$, we obtain $\psi^{\alpha\beta(-1,2,-1)}\oplus 2\,\psi^{\alpha\beta(-1,-1,2)}$. Here, we used the traceless family numbers as in Sec. \ref{sec:RK}.}
 \dis{
 2\psi^{\alpha\beta(-1,2,-1)}\oplus \psi^{\alpha\beta(-1,-1,2)}.\label{eq:ThreeTens}
 }
  
 \subsection{Changing chirality}\label{subsec:Chirality}
\begin{figure}[!t]
\hskip 0.01cm \includegraphics[width=0.8\textwidth]{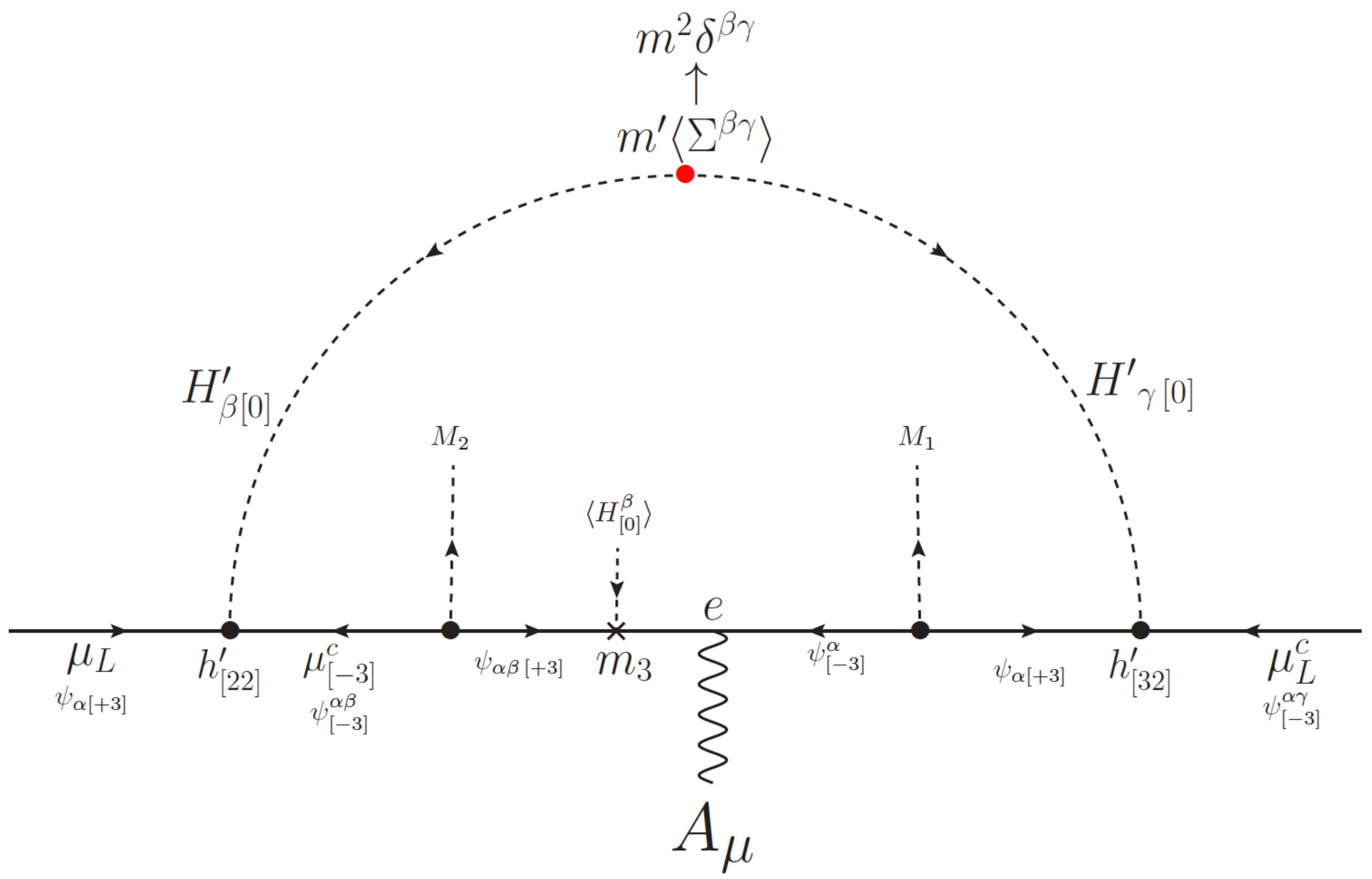}
\caption{A BSM contribution to $(g-2)_\mu$. $m_3$ can be placed at the LHS (as here) and RHS of the $A_\mu$ vertex, and $\langle H^{\beta}\rangle $ is a VEV of the SM Higgs doublet giving mass to $\mu$. $\Sigma^{\beta\gamma}$ is a SU(5) singlet field $\propto \delta^{\beta\gamma}$.}\label{fig:gm2}
\end{figure}

 \noindent
The BSM contributions to the magnetic moments need a change of chirality \cite{Kim76}. The effective interaction for changing chirality of leptons is through $\psi_{\alpha}\leftrightarrow \psi^{\alpha\beta}$: $f^{(e)}\psi^{\alpha\beta}\psi_{\alpha} H_{\beta}^{(d)} $. In the SM, it gives masses to $\Qem=-1$ leptons. If we consider only one Higgs doublet, there is no BSM contribution to the magnetic moments. We need more Higgs doublets. Let us denote this extra (inert) Higgs doublet without VEV as $H_{\beta}^{'}$, and its coupling to the muon family  
\dis{
h'_\mu \psi^{\alpha\beta(-1,2,-1)}\psi_{\alpha}^{(1,-2,1)} H'_{\beta} \label{eq:BSMcoup}
}
introduces the BSM contribution to the magnetic moments. 
Note that $\psi^{\alpha\beta(0,1,0)}_L$ can be represented as  $\psi_{\alpha\beta(0,-1,0),R}$ moving in the opposite direction. We used the muon quantum number to have additional contribution to $(g-2)_\mu$. In Eq. (\ref{eq:light}), $\mu$ is in $\psi_{\alpha}^{(0,-1,0)}$ and $\mu^c$ is in $\psi^{\alpha\beta (0,1,0) }\equiv \psi_{\alpha\beta (0,-1,0)R}$, \ie $\mu_L$ has muon number +1 and $\mu_R$ also has muon number +1.  From Fig. \ref{fig:gm2}, we estimate 
\dis{
a_\mu'=\frac{(g-2)_\mu}{2}\propto e h_{[22]}' h_{[23]}'.
} 
Here,  $m, M_1$ and $M_2$ can be superheavy but  $m_3$ must be smaller than or at the electroweak scale.

 \bigskip
 \noindent
The LHS and RHS contributions of Fig. 1 sandwiched between $\bar{u}(p') $ and $ u(p)$ are   
 \dis{
LHS &\propto m^2  m_3 e h_{[22]}' h_{[23]}'\int\frac{d^4 k}{(2\pi)^4}    \frac{(k'\hskip-0.26cm \slash+M_1)\gamma_\mu (k\hskip-0.17cm \slash+M_1)(k\hskip-0.17cm \slash+M_2)}{({k'}^2-M_1^2)({k}^2-M_1^2)({k}^2-M_2^2)\Big({(p-k)}^2-m^2\Big)^2},
 }
 and similarly for the RHS. Here we assumed $m^2$ is the $H'_{\alpha[0]}$ mass and treated $m_3$ as the mass of  $\psi_{\alpha\beta[+3]}\oplus \psi^\alpha_{[-3]}$, and
 \dis{
 p=\textrm{momentum of }\mu_L , p'=\textrm{momentum of }\mu_L^c ,q=p'-p, k'=k+q.
 }
 Thus, from Fig. 1  the anomalous magnetic moment is estimated as
\dis{
F_2(0)=\frac{ e h_{[22]}' h_{[23]}'m^2 m_3 }{8\pi^2 (M_2^2-M_1^2)}\int_0^1 dx\int _0^x dy~ (y-x)\Big\{ \cdots\Big\} .
}
where $\cdots$ is 
\dis{
\Big\{ \cdots\Big\} =&\left(M_1M_2 (x-y-1)-(1-3y)(x-y)m^2 \right)
\left(\frac{1}{\Delta_1}+\frac{1}{{\Delta_2'}}-\frac{1}{\Delta_2}-\frac{1}{{\Delta_1'}}\right)\\
&+\frac{(3y-2)(1-x)M_1^2+y(3y-1)M_2^2}{\Delta_1} \color{black}- \frac{y(3y-1) M_1^2+(x-1)(2-3y)M_21^2}{\Delta_2}\\
&+\frac{[2(x+y-1)-3y(x-y)]M_2^2}{{\Delta_2'}}- \frac{[2(x+y-1)-3y(x-y)] M_1^2}{{\Delta_1'}}\,.\label{eq:cdots}
}
The calculation through the Feynman parametrization is sketched in the Appendix.

\bigskip
\noindent
For $M_1,M_2\ll m$, ${\Delta_1}\simeq {\Delta_2}\simeq {\Delta_1'}\simeq {\Delta_2'}\simeq (x-y) m^2$.
In this case, the integral is simplified to
\begin{eqnarray}
F_2(0)&=&\frac{ e h_{[22]}' h_{[23]}'  m_3 m_\mu}{8\pi^2 \,m^2}
\int_0^1 dx\int _0^x dy\,   \frac{2y(1-3y)}{x-y}\nonumber \\
&=&-\frac{ e h_{[22]}' h_{[23]}'  m_3 m_\mu}{8\pi^2 \,m^2}\left(\frac{13}{6}+\ln\epsilon \right)
 \label{eq:gm2muprime}
\end{eqnarray}
where $\epsilon=\frac{\{M_1^2,M_2^2\}}{m^2}$.

\bigskip
\noindent
For the other extreme $M_1,M_2\gg m$, ${\Delta_1}\simeq (1-x) M_1^2+(x-y)M_2^2, {\Delta_2}\simeq yM_1^2+(1-x)M_2^2, {\Delta_1'}\simeq (1-x+y)M_1^2, {\Delta_2'}\simeq  (1-x+y)M_2^2$.
In this case, the integral is simplified to give 
\dis{
F_2(0)=\frac{ e h_{[22]}' h_{[23]}'  m_3 m_\mu}{16\pi^2\,M_1^2 }\left(\frac{1+\xi+\xi^2}{\zeta^2\xi^3}\right).
}
where $\zeta=M_1/m$ and $\xi=M_2/M_1$. For the $R_K$ phenomenology, we will need $\frac{m}{\sqrt{ h_{[22]}' h_{[23]}'  }}\approx O( 0.7\times 10^6\,\gev)$. So, suppose $m=10^{6\,}\gev, M_2\lesssim M_1, M_1=10^4\,\gev, m_3\approx 200\,\gev$, and $h_{[22]}' \simeq h_{[23]}'\approx 1$, which gives $F_2(0)\approx 1.21\times 10^{-9}$. 

\bigskip
\noindent
The BNL value of $(g-2)_\mu$ minus the SM prediction is \cite{MarcianoPDG},
\dis{
\Delta a_\mu=a_\mu^{\rm exp}- a_\mu^{\rm SM}
=261\,(63)\,(48)\times 10^{-11}.\label{eq:BNLmu}
}
Thus, there is some region of the parameter space pulling the $(g-2)_\mu$ of the SM value to the BNL value with $m=O(10^6)\,\gev$, with $m_3$ at the electroweak scale with heavy $m,M_1,$ and $M_2$. 

\color{black}

\bigskip
\noindent
If we consider the symmetry \Umutau, the calculation is the same.

 \subsection{Models for $R_{K,K^*}$}
 \noindent
For the fGUT interaction discussed in Subsec. \ref{subsec:Chirality}, let us consider what can be its effect to $R_H$ with  the symmetry \Umutau. 
From Eq. (\ref{eq:ThreeTens}), the $L_\mu-L_\tau$ quantum numbers of three families in the GG model are
\dis{
 \psi^{\alpha\beta }_{[-3]}  \oplus    2\,\psi^{\alpha\beta}_{[+3]} \oplus   2\,  \psi_{\alpha[-3]}\oplus   \psi_{\alpha[+3]}
}
and the BSM Higgs field $H'$ can be one from $(\psi^{\alpha}_{[0]}\oplus \psi_{\alpha[0]})$'s.  To have the coupling (\ref{eq:BSMcoup}), $H'$ must be neutral in $L_\mu-L_\tau$. 
Thus, the $H'$ couplings takes the form,
\dis{
h'_{[IJ]} \psi^{\alpha\beta}_{[I]}\psi_{\alpha[J]} H'_{\beta[0]} \label{eq:BSMflco}
}
 From (\ref{eq:BSMflco}), we note that $b_L$ can decay to $s_R$,
\begin{eqnarray}
b_L({\rm from} ~ \psi^{\alpha\beta}_{[+3]})\to s_L^c({\rm from}~\psi_{\alpha [+3]})+{\rm neutral~Higgs~}({\rm from}~H'_{[0]}~\rm possible).\label{eq:allowed0}
\end{eqnarray}
In the next section, we will use the following interaction
\dis{
 h'_{[23]} \overline{s}_R b_L H{'}^{0}+h'_{[22]}\overline{\mu}_R \mu_L H{'}^{0}+{\rm h.c.}\label{eq:allowRK}
}
where $h'_{[IJ]}$ couplings are set to real values by absorbing their phases to quark or lepton fields.

\section{Phenomenology of $L_{\mu}-L_{\tau}$ for quarks via ${\bf f}$GUT}\label{sec:RK}

\noindent
From the allowed coupling (\ref{eq:allowRK}) with the $L_\mu-L_\tau$ conservation, we will estimate $R_K$. The following contribution to $R_K$ exists by the BSM neutral field $H{'}^{0}$,
\dis{
\frac{ h'_{[23]} h'_{[22]}}{m_0^2}\,\bar{\mu}_L\mu_R  \bar{s}_R b_L +{\rm h.c.}\label{eq:BSMtoRK}
}
We choose traceless combinations for the flavor indices. Thus, note that   $ \psi_{\alpha (-1,2,-1)}$ houses the muon doublet:    $ \psi_{\alpha (-1,2,-1)} =\psi_{\alpha [3]} $ where the number in the bracket [~] is the $L_\mu-L_\tau$ quantum number. The tau doublet belongs to   $ \psi_{\alpha (-1,-1,2)} =\psi_{\alpha [-3]} $.
The quark doublet of the 2nd family is in $\psi^{\alpha\beta(-1,2,-1)} =\psi^{\alpha\beta}_{(1,-2,1)} =\psi^{\alpha\beta}_{[-3]}$ which is in the L-handed notation. In terms of the R-hand notation,  the $L_\mu-L_\tau$ quantum number is [3]. The quark doublet of the 3rd family is in $\psi^{\alpha\beta(-1,-1,2)} =\psi^{\alpha\beta}_{(1,1,-2)} =\psi^{\alpha\beta}_{[3]}$ which houses $b_L$. $\mu_L^c$ is contained in $\psi^{\alpha\beta(-1,2,-1)}=\psi^{\alpha\beta}_{(1,-2,1)}=\psi^{\alpha\beta}_{[-3]}$. Thus, the fields participating in the $b$ decay via the BSM field  $H{'}^{0}$ are
\dis{
&\mu_L:  \psi_{\alpha [+3]}  \\
&b_L: \psi^{\alpha\beta}_{[+3]}\\
&\mu_R: \psi^{\alpha\beta}_{[+3]}\\
&s_R: \psi^{\alpha\beta}_{[+3]}
} 
Thus, Eq. (\ref{eq:BSMtoRK}) preserves the U(1)$_{\mu-\tau}$ gauge symmetry. 
Note also that  the combination $h'_{[23]} h'_{[22]}$  appeared in the BSM contribution to $(g-2)_\mu$.  

\bigskip
\noindent
The lepton flavor universality from  one-loop generated coupling in the SM  is given by \cite{Bobeth11},
\dis{
\frac{G_F\alpha_{\rm em} V_{tb}V^*_{ts}\,C_9}{\sqrt2 \pi} \overline{\ell}\gamma^\mu(1-\gamma_5) \ell\,\overline{s}_L\gamma_\mu b_L\label{eq:SMHam}
}
where $\ell=e,\mu,\tau$ and $C_9=(-\frac12+\sin^2\theta_W)(\frac12-\frac23\sin^2\theta_W)$ times some factor arising from hadronic states.  

\bigskip
\noindent
The $B^+$ meson decay rate with lifetime $1.638(1.000 \pm 0.003)\times 10^{-12}$ s is \cite{HFLAV19},
\dis{
\Gamma_{B^+}\simeq 4.02\times 10^{-13} \gev
}
while the branching ratio for  $\bar{B}\to \bar{K}\ell^+\ell^-$ (for $\ell=e,\mu$) is  \cite{Bobeth11},
\dis{
{\cal B}(\bar{B}\to \bar{K}\ell^+\ell^-)=1.04\times 10^{-7}.
}

\begin{figure}[!t]
\hskip 0.01cm \includegraphics[width=0.9\textwidth]{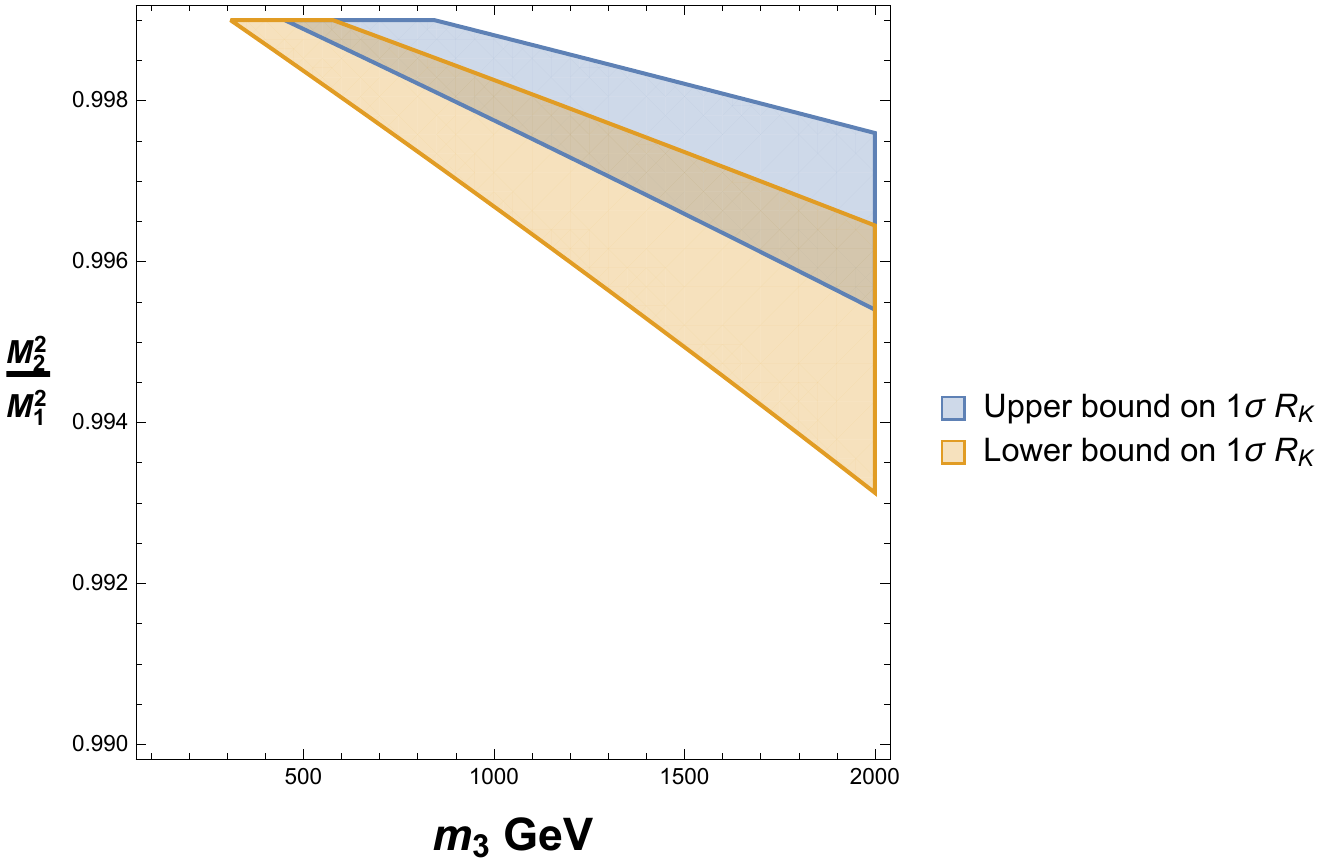}
\caption{Allowed region of parameter space ($ \frac{M^2_2}{M^2_1}, m_3$) from  $R_K$ and $(g-2)_\mu$ bounds for $M_1^2=10 m_0^2$.
}\label{fig3}
\end{figure}

 \color{black}
\bigskip
\noindent
The family dependence is studied by a double ratio  \cite{LHCbRK14,LHCbRK19},
\dis{
R_K &=\frac{{{\cal B}(B^+\to K^+\mu^+\mu^-)}}{{\cal B}(B^+\to K^+J/\psi(\to \mu^+\mu^-))} \Big/\frac{{{\cal B}(B^+\to K^+e^+e^-)}}{{\cal B}(B^+\to K^+J/\psi(\to e^+e^-))}\,.  
}
The first report on  the family dependence is the Run1 result of LHCb, $R_K=0.745^{+0.090}_{-0.074}\pm 0.036$ 
in the $q^2$ interval of $q^2=1.1-6\,\gev^2$ \cite{LHCbRK14}, which gives a   2.6 $\sigma$ level anomaly,  away from the SM prediction. The recent result from LHCb is $R_K=0.846^{+0.060+0.016}_{-0.054-0.014}$ \cite{LHCbRK19}, which is a 2.5 $\sigma$ anomaly. However, the recent Belle report is consistent with the SM but with larger error bars,  $R_K=0.98^{+0.27}_{-0.23}\pm 0.06$ \cite{Belle19}.  We will use the Run1 result of LHCb since it covers a wide range of $q^2$,
\dis{
R_K  =0.745^{+0.090}_{-0.074}\pm 0.036.\label{eq:RKdata}
}

\bigskip
\noindent
There exists a claim that  new physics by scalar mediation cannot explain the $R_K$ phenomenology \cite{Alonso14} (see also \cite{Hiller14}). But, their assumptions deriving this conclusion do not include our scenario. Firstly, they assumed only the SM gauge group while our symmetry below the cutoff scale $\Lambda$ is the SM gauge group times  \Umutau. Second, the constraint, for example Eq. (15) of   \cite{Alonso14}, is from purely leptonic data. But, our interaction Eq. (\ref{eq:BSMtoRK}) is not related to the SM interaction and hence the parameters in  Eq. (\ref{eq:BSMtoRK}) are restricted only by the $(g-2)_\mu$ phenomenology, whose allowed region will be given together with the $R_K$ bound.
   
 \bigskip
 \noindent
Incorporating the new operator $C_{bs\mu\mu}\,\bar{\mu} \mu\,\bar{s} P_{R(L)}b$, $R_K$ is given by
\dis{R_K=\frac{\Gamma_{\mu}}{\Gamma_e}=1-\frac{\Gamma_{\mu}^{\rm NP}}{\Gamma^{\rm SM}_{e}}, 
}
where, 
\dis{\Gamma^{\rm SM}_e\simeq \frac{1}{3}\int^{q_{max}}_{q_{min}}
\Gamma_0 \beta_e \lambda^{3/2}\xi_P^2(q^2)(C^2_{9~eff}+C^2_{10}) dq^2,\label{eq:SM}
}
\dis{\Gamma^{\rm NP}_{\mu}\simeq \int^{q_{max}}_{q_{min}}
\Gamma_0 \beta_\mu^3 \lambda^{1/2}\xi_P^2(q^2)|F_s|^2|C_s|^2dq^2,\label{eq:NP}
}
with
\dis{\Gamma_0=\frac{G^2_F \alpha_{em}^2|V_{tb}V^{\ast}_{ts}|^2}{512 \pi^5 M^3_B},~~~\beta_l=\sqrt{1-\frac{4m^2_l}{q^2}}\simeq 1.
}
\dis{\lambda=M^4_B+M^4_K+q^4-2(M^2_B M^2_K+(M^2_B+M^2_K)q^2,
}
\dis{\xi_P(q^2)\simeq \frac{  0.327
}{(1-\frac{q^2}{M^2_B})^2}, ~~~F_S=\frac{M_B^2-M^2_K}{m_b-m_s}\frac{(M^2_B+M^2_K-q^2)}{M^2_B}.
}
 
 \bigskip
 \noindent
In our model, the contribution of  $C_{bs\mu\mu}$ relative to the one in the SM  is given by
\dis{C^2_s\equiv \left(\frac{\sqrt{2}\pi}{G_F\alpha_{em}V_{tb}V^{\ast}_{ts}}\right)^2\left(\frac{ h'_{[23]} h'_{[22]}}{m_0^2}\right)^2.
}
 
 \bigskip
 \noindent
From the experimental results, Eq.(\ref{eq:RKdata}), we can determine the values of $\frac{ h'_{[23]} h'_{[22]}}{m_0^2}$.
For example, taking the central value of Eq.(\ref{eq:RKdata}),  we obtain
\dis{
\left| \frac{\Gamma_{\mu}^{\rm NP}}{\Gamma^{\rm SM}_e}\right|=1-0.745\simeq 0.03|C_s|^2,
} 
where we performed the $q^2$ integration in Eqs. (\ref{eq:SM}) and (\ref{eq:NP}). With the known SM parameters,   $M_B, M_K, m_b, m_s=5.2795, 0.89594, 4.8, 0.101$ in GeV units, respectively,  $|V_{tb}|=0.999097$ and $|V^{\ast}_{ts}|=0.04156$, we obtain   $\frac{ h'_{[23]} h'_{[22]}}{m_0^2}\simeq 2.5\times 10^{-9}~{\rm GeV}^{-2}$. 
Determining $\frac{ h'_{[23]} h'_{[22]}}{m_0^2}$ in the way, we can compare the  $(g-2)_\mu$ shift by Fig. \ref{fig:gm2} with  the measured value at the BNL. We note that there are more  unknown parameters for the expression on the muon $(g-2)_\mu$: $m_3,M_1,M_2$ for a given
 $\frac{ h'_{[23]} h'_{[22]}}{m_0^2}$.

\bigskip
\noindent
To glimpse a behavior for $(g-2)_\mu$,  we choose $M^2_1/m^2_0$  to be 10 and look for the allowed region of $\frac{ h'_{[23]} h'_{[22]}}{m_0^2}$ up to 1 $\sigma$ region of experimental result of $R_K$.
Then,  $m_3$, and $ \frac{M^2_2}{M^2_1}$ are the only unknown parameters for the $(g-2)_\mu$ expression.
Imposing the experimental result, Eq(\ref{eq:BNLmu}), we can get the allowed region of parameters space
$(\frac{M^2_2}{M^2_1}, m_3)$, which is shown in Fig. \ref{fig3}.
In Fig. \ref{fig3}, the blue and  orange regions correspond to the alowed regions
of parameter space obtained by taking $R_K$ to be the $1~\sigma$ upper limit of $ R_K$,
and the $1~\sigma$ lower limit of $ R_K$.
\color{black}

\section{Conclusion} 

\noindent
In a family grand unification model, we related the BNL anomaly on the muon anomalous magnetic moment and the LHC anomaly on $R_K$ 
via the symmetry $L_\mu-L_\tau$. As a fGUT grand unification example, we used  Georgi's SU(11).

\acknowledgments{\noindent JEK  and SKK are supported in part  by the National Research Foundation 
(NRF) grants  NRF-2018R1A2A3074631 and   NRF-2019R1A2C1088953, respectively.}
\vskip 1cm

\centerline{\bf Appendix}
  \vskip 0.5cm
 The useful Feynman parameter integration is
 \dis{
 \frac{1}{ab^2cd}=-\frac{\partial}{\partial b} \left(\frac{1}{abcd} \right)=-\frac{1}{(d-c)}\frac{\partial}{\partial b} \left(\frac{1}{abc} -\frac{1}{abd} \right)
 }
  where
 \begin{eqnarray}
\frac{1}{abc}=-2!\int_0^1dx\int_0^x dy\, \frac{1}{\left[a + (b-a)x +(c-b)y\right]^3}\label{eq:D1}\\
\frac{1}{abd}=-2!\int_0^1dx\int_0^x dy\, \frac{1}{\left[a + (b-a)x +(d-b)y\right]^3} \label{eq:D2}
 \end{eqnarray}
 Let us choose, for the (LHS) of Fig. 1,
 \dis{
 &a={k'}^2-M_1^2,b={(p-k)}^2-m^2,c={k}^2-M_2^2,d={k}^2-M_1^2,\\
& \ell=k+ q(1-x)+p(y-x) =k'-qx+p(y-x).
 }
 Equations (\ref{eq:D1}) and (\ref{eq:D2}) have the following denominators
 where
 \dis{
 D_L=a+(b-a)x +(c-b)y=\ell^2-\Delta_1\\
 D_L'=a+(b-a)x +(d-b)y=\ell^2-\Delta_1'
 }
  where, neglecting O($m_\mu^2$), 
 \dis{
& \Delta_1= M_1^2(1-x) +M_2^2y+ m^2(x-y),\\
& \Delta_1'=M_1^2(1-x+y) + m^2(x-y).
 }
 A similar consideration for the (RHS) of Fig. 1 leads to, for $D_R$ and $D_R'$,
  \dis{
& \Delta_2= M_2^2(1-x) +M_1^2y+ m^2(x-y),\\
& \Delta_2'=M_2^2(1-x+y) + m^2(x-y).
 }
   

  \end{document}